\documentclass[preprint,showpacs,preprintnumbers,amsmath,amssymb]{revtex4}

\usepackage{graphicx}
\usepackage{subfigure}
\usepackage{epsfig}
\usepackage{graphics}
\usepackage{picins}

\usepackage{amssymb}
\usepackage[dvips]{color}
\begin{document}
\title{Stability and magnetic properties of Mn-substituted ScN from first principles}
\author{Abdesalem ~Houari$^{a)}$, Samir~F.~Matar$^{b)}$\footnote{Corresponding author (matar@icmcb-bordeaux.cnrs.fr)},
Mohamed A. Belkhir$^{a)}$}
\affiliation{$^{a)}$Laboratoire de Physique Th\'eorique, D\'epartement de Physique, Universit\'e de Bejaia. 06000 Bejaia, Algeria. \\
$^{b)}$ICMCB, CNRS, Universit\'e Bordeaux 1. \\ 87 avenue Dr A Schweitzer. 33600 Pessac, France.}
\date{\today}

\begin{abstract}
We present a spin density functional theory (DFT) study for semiconducting ScN and Mn-substituted ScN. 
Their structural and  magnetic properties have been investigated using the all electrons augmented spherical wave method (ASW) with a generalized gradient GGA functional for treating the effects of exchange and correlation.
Band structure calculations show that ScN is semiconductor with a narrow indirect band gap $\Gamma$-X of 0.54 eV. 
The total-energy versus volume calculations show that ternary Sc$_{0.75}$Mn$_{0.25}$N nitride is more stable in face centered  tetragonal-rocksalt (fct-rocksalt) structure, found experimentally, than the cubic rocksalt one. Spin polarized results, at  theoretical equilibrium, indicate that the ground state of Sc$_{0.75}$Mn$_{0.25}$N is ferromagnetic with a high moment at  Mn-atom (3.44$\mu_B$), and zero moment on Sc and N ones. The magnetovolume effects of Mn-substitution in ScN lattice are discussed. The electronic structures analyzed from site/spin projected density of states and chemical bonding, for both the mononitride and the  ternary alloy, are reported. A discussion of the structural and magnetic properties of Sc$_{0.75}$Mn$_{0.25}$N is given with a comparison to ScN, in order to get insights of the Mn-substitution effects. 
\end{abstract}
\maketitle

\section{Introduction}
With the emergence of the spintronic field, the search of new dilute magnetic semiconductors (DMS) and materials for spin injection is of great interest. Currently, the doped III-V semiconducting nitrides, such as GaN and InN, by magnetic transition metal atoms (Cr, Mn, Fe ...etc) constitute promising candidates due to their predicted high (room) Curie temperature \cite{Dietl}.  However, major experimental limitations exist in this domain, such as low solubility of the magnetic dopant for DMS which is related to the different crystal strcutures (between the host and the dopant) or the mismatch in resistivity for spin injection when a metallic ferromagnet and a semiconductor are joined together ~\cite{Herwad,Schmidt}. For the cited applications, the late transition mononitrides, which have a metallic character, i.e. MnN, FeN and CoN, have recently attracted much interest (see~\cite{Houari} and reference therein).  
On the other side, the early ScN, TiN and VN nitrides, well known to crystalize in rocksalt (RS) strcuture, were largely 
explored for their thermal and mechanical properties \cite{Oyama}. However, recent progress in synthesis of stoichiometric pure ScN have reopened the debate about this nitride, especially for a possible use in the applications cited above. Early experiments reported ScN to be a semimetal \cite{Travag}, confirming the results \cite{Monnier} of density-functional theory (DFT) calculations within the local density approximation (LDA)~\cite{Kohn,Sham}. More recently, however, other experimental studies, by means of optical absorption, have found it as an indirect band gap semiconductor ~\cite{Mousta,Albrithe,Smith,Albrithen}  where a value of 0.9 eV was given by Al-Brithen {\it et al} \cite{Albrithen}.  After that and in order to overcome the well known drawbacks of the LDA in (under)estimating the gap, several other theoretical  investigations, using more accurate approximations going beyond the local density one, have been undertaken  to explore ScN electronic structure ~\cite{Lamb,Gall,Stampfl,Simunek}. Based on ideas of GW approwimation \cite{Hedin}, Lambrecht~\cite{Lamb} found an indirect $\Gamma$-{\it X} gap for both GdN and ScN, with
 a value of 0.9 eV for the latter. A full-potential calculation, using the screened-exchange local density (sX-LDA),
 has shown that bulk ScN \cite{Stampfl}, as well as in thin films \cite{Stamp}, is also semiconducting with
an indirect gap of 1.58 eV in the bulk. However no gap value was provided from the surface study; nevertheless from the band  structure plots, the gap value seems larger than 1.5 eV \cite{Stamp}. Very recently, Qteish {\it et al} \cite{Qteish} have theoretically invistigated ScN with pseudopotential calculations, using different approximations for the exchange-correlation  effects. With more recent quasiparticle approximation (OEP-cLDA-GW), see \cite{Qteish} and therein cited references for more details, the authors again arrived at the experimental 0.9 eV gap value. It is important to note that for some authors \cite{Stampfl,Qteish}, the use of the generalized gradient approximation (GGA), with different paramerizations, \cite{GGA-P,PBE} respectively, has led to a semimetalic character. From the cited works, it is clear that the use of an accurate approximation of the exchange-correlations effects is crucial in predicting the true energy gap, in as far as this is exactly the physical quantity which is being investigated.
 With a lattice parameter of {\it a} = 4.5~ \AA ~\cite{Park,Gubanov,Schilf}, ScN mononitride is good lattice matched to the wide band-gap GaN semiconductor. Furthermore, its cubic rocksalt structure, like MnN (distorded-rocksalt one) \cite{Suzuki} or perhaps FeN \cite{Houari}, make it very attractive as a possible magnetic semiconductor when doped or  substituted by magnetic atoms but. Unfortunately, only few reports are available in the literature. To the best of our knowledge only one recent exprimental study on Sc$_{(1-x)}$Mn$_{x}$N was carried out.
 Al-Brithen {\it et al} \cite{Albri} have incorporated Mn to ScN and obtained Sc$_{(1-x)}$Mn$_{x}$N nitrides solid solution 
 up to x=0.25 with a face-centred tetragonal rocksalt-type structure. They found that 'ScMnN' lattice constants {\it a} and  {\it c} follow linear Vegard's law going from pure ScN to pure $\theta$-phase MnN, but no magnetic data were given. On the  theoretical side, also the only study of Mn-doped ScN  (calculations of 64 atom supercell of rocksalt ScN doped with a single Mn atom on a Sc site) have shown that the Mn atom induces a localized state in the middle of the band-gap with a  $2\mu_B$-$3\mu_B$ magnetic moment \cite{Herwad}.  
 
Due to the lack of data on manganese doped ScN on one hand and in order to provide an improved theoretical insight on the other hand, we carried out a comprehensive spin density functional calculation. In view of the large number of works on pure ScN , its study is briefly included here for sake of comparison and completeness. From this, the main purpose of the present work is the investigation of the electronic and magnetic structures of Sc$_{0.75}$Mn$_{0.25}$N. Both a cubic rocksalt (RS) and a face centered tetragonal rocksalt (fct-RS type) structures were considered. 
The structural, electronic and magnetic properties are thus studied by analyzing the calculated electronic structure, total energies and crystal field influence. Our  paper will be organized as follows: A description of the computational details is given in section 2. Our results for the calculated total energy and electronic structure of ScN are presented in section 3. Section 4 which is devoted to the Sc$_{0.75}$Mn$_{0.25}$N alloy is divided in two parts: in the first one the magnetic properties are discussed within the framework of the Stoner theory of band ferromagnetism, while in the second one the electronic structure is commented  via partial density of states (DOS) and chemical bonding analysis. Finally a conclusion is given in the last section.  

\section{Computational details}

The calculations of the present study are performed in the framework of the density functional theory ~\cite{Kohn,Sham} using the generalized gradient approximation (GGA) with the Perdew, Burke and Ernzerhof PBE parameterization for  the effects of exchange and correlation ~\cite{PBE}. This parameterization scheme was preferred over the LDA in view of its drawbacks in estimating total energies and lattice constants magnitudes.

The scalar-relativistic augmented spherical wave method (ASW) ~\cite{Kubler,Eyert} based on the  atomic sphere approximation (ASA) is mainly used. In this method, the wave function is expanded in atom-centered augmented spherical waves, which are Hankel functions and numerical solutions
 of Schr\"odinger's equation, respectively outside and inside the so-called augmentation spheres. In order to optimize the  basis set, additional augmented spherical waves were placed at carefully selected interstitial sites. The choice of these sites as well as the augmentation radii were automatically determined using the sphere-geometry optimization SGO  algorithm ~\cite{Sgo}. Our choice of radii minimizing the overlap between ASA spheres was as follows: r$_{Mn}$=1.34 \AA,  r$_{Sc}$=1.36 \AA, r$_N$= 1.14 \AA ~and r$_{ES}$ at ($\frac{1}{4}$,$\frac{1}{4}$,$\frac{1}{4}$)=0.85 \AA; ES designate empty spheres with zero atomic number, introduced within the SGO algorithm. The basis set consisted of Mn(4s, 4p, 3d), Sc(4s, 4p, 3d) and N(2s, 2p) valence states; ES were given basis set similar to N for allowing charge transfer. The Brillouin zone (BZ) integrations were performed with an increasing number of $k$-points (16x16x16) in order to ensure convergence of the results with respect to the $k$-space grid. The convergence criterion is fixed to 0.001 mRy in the self-consistent procedure  and charge difference $\Delta~Q$=10$^{-4}$ between two successive iterations. In order to establish a reference for the spin-polarized calculations we started with a set of spin-degenerate calculations.  Such a configuration is however not relevant to a paramagnet which would only be simulated by a huge supercell entering random spin orientations over the different magnetic sites or alternatively by calling for disordered local moment approach, which is based on coherent potential CPA approximation ~\cite{cpa} or for the LDA+DMFT approach ~\cite{ldadmft}. It allows assigning a role to the orbitals responsible of the magnetic instability toward spin polarization in a mean field analysis using the Stoner theory of band ferromagnetism ~\cite{Stoner}.

Further we examine the bonding using the  "Energy of Covalent Bond" (ECOV) criterion \cite{ECOV} which merges both overlap S$_{ij}$ and Hamiltonian H$_{ij}$ (i, j being two chemical species) analyses to extract chemical bonding characteristics. In ECOV plots along coordinate y-axis,  a positive magnitude points to an antibonding character (destabilizing) , whereas a negative value assigns a bonding character (stabilizing) of the bond. 

For sake of completeness, we used two other complementary methods by providing the electronic density of states of ScN and an optimized geometry for  Sc$_3$MnN$_4$ (i.e., Sc$_{0.75}$Mn$_{0.25}$N) lattice starting from experimental determinations using respectively a full potential LAPW method \cite{wien} and a pseudo-potential method \cite{vasp}. 
  
\section{Band structure of scandium nitride}

\begin{figure}[h]
\begin{center}
\includegraphics[width=0.4\linewidth]{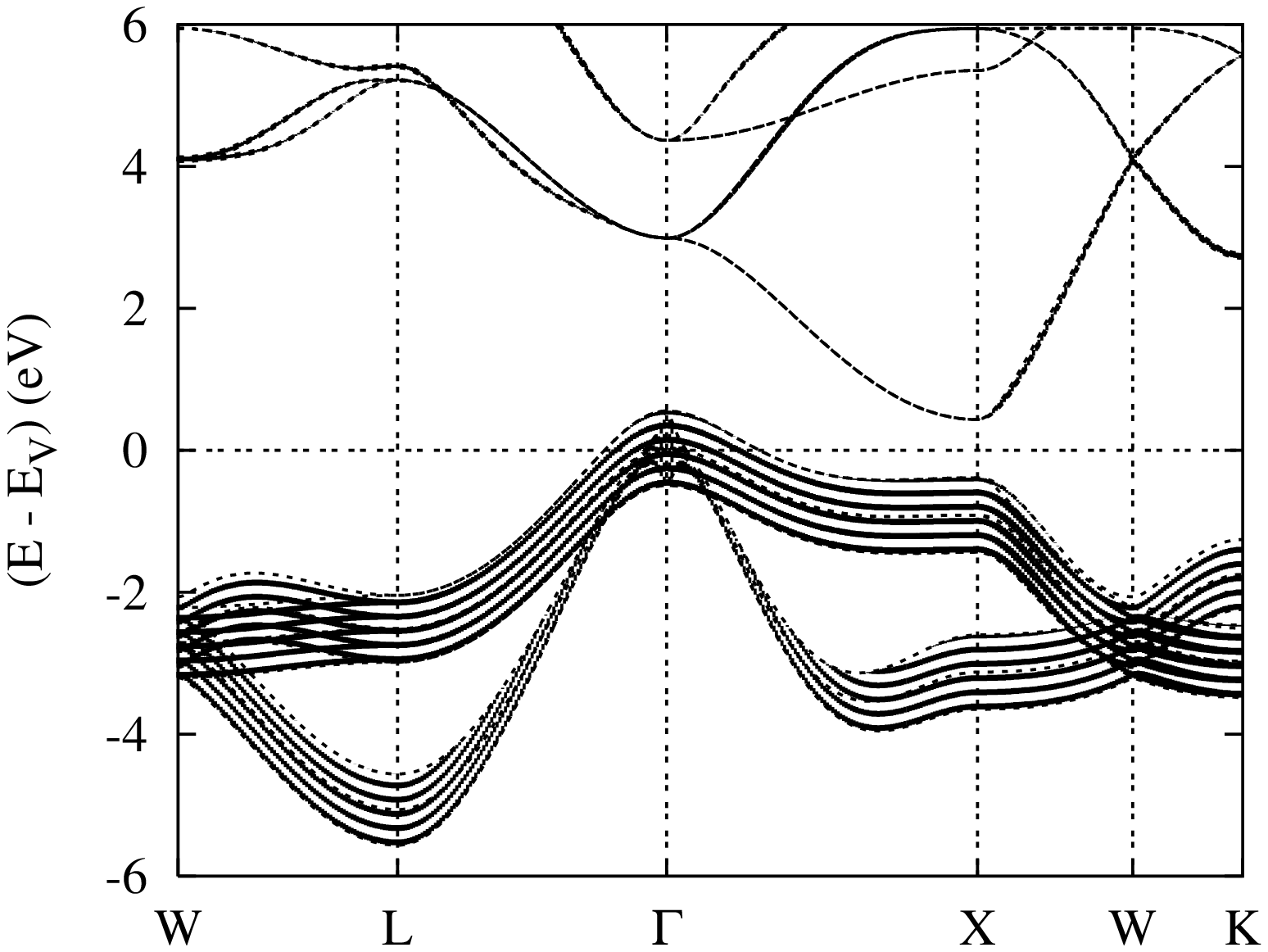}{  a)}
\includegraphics[width=0.4\linewidth]{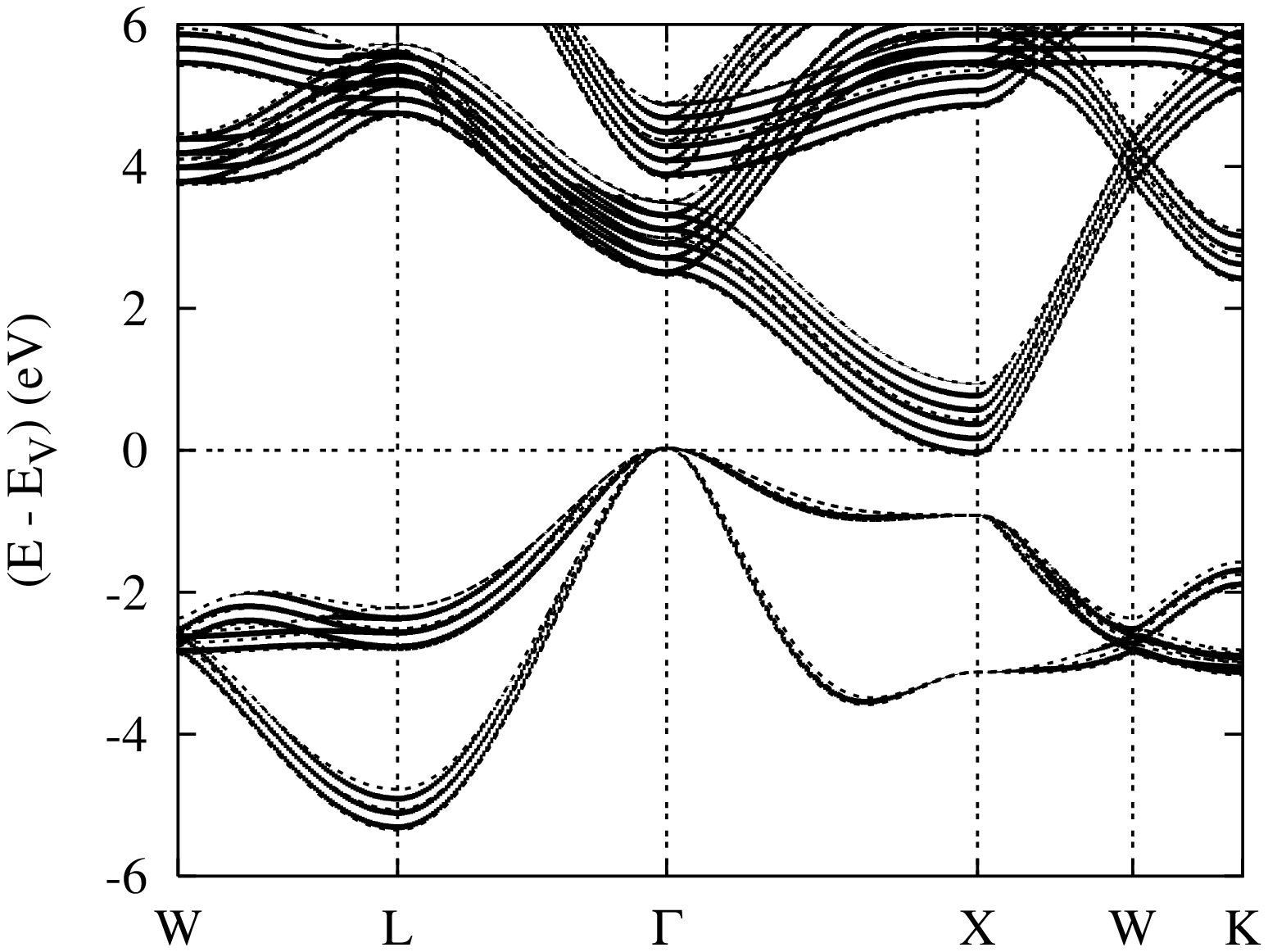}{  b)}
\caption{Band structure of ScN at theoretical equilibrim volume: a) Fat bands weighted with N 2p states, b) Fat bands weighted with Sc 3d states.}
\label{fig1}
\end{center}
\end{figure}

The obtained numerical results for the electronic band structure of ScN are summarized in table I together with some data from literature. Our calculated equilibrium lattice constant value of {\it a} = 4.5 \AA ~is found  in  agreement with both experimental ~\cite{Park,Gubanov,Schilf} and theoretical \cite{Stampfl} works. The band structure is shown in figs. \ref{fig1}. From fig. \ref{fig1}-a, the valence band (VB) is dominated by nitrogen and one can count three $p$-like bands. This is illusttrated by the fat bands character weighted with N-2p states. An indirect $\Gamma$-X character of the band gap ($\sim$ 0.54 eV) is in agreement with recent experimental and theoretical reports but its magnitude is lower (see \cite{Lamb} and references therein). This reduced gap magnitude is due to the large dispersion of the bands along the $\Gamma$-X-W direction which is generally exaggerated within exchange-correlation functionals based on the electron gas approximation used here. In the conduction band (CB), within an $O_h$-like crystal field, one can actually count 3 bands for  the  $t_{2g}$ manifold followed at higher energy by two  $e_g$ bands. The character of these bands is of Sc- 3d nature as shown by the fat bands as illustrated in fig. 1-b. However one can notice a small Sc-3d weight within the VB which is a signature of the covalent bonding between Sc and N. 

\begin{figure}[h]
\begin{center}
\includegraphics[width=0.4\linewidth]{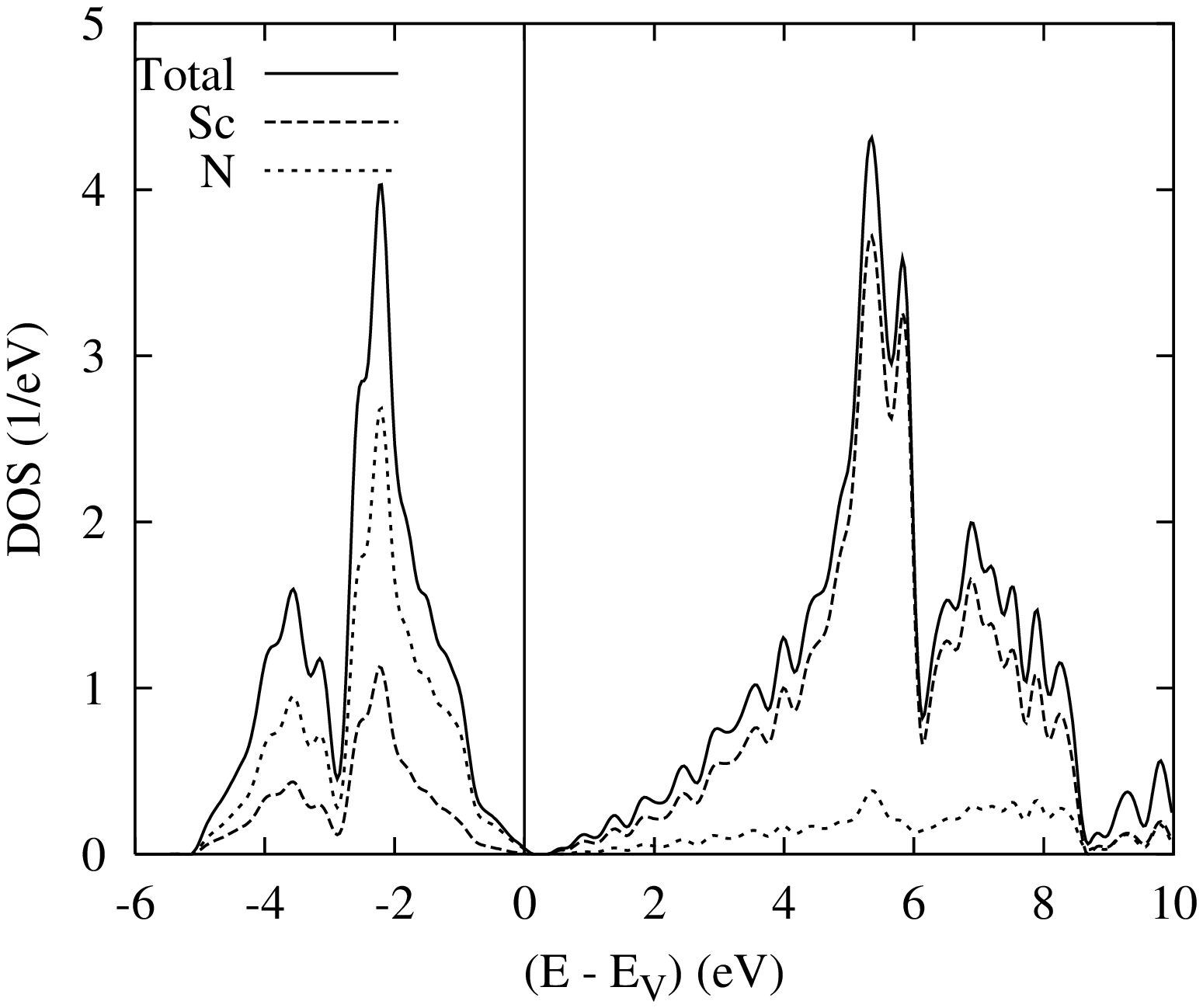}{a)}
\includegraphics[width=0.5\linewidth]{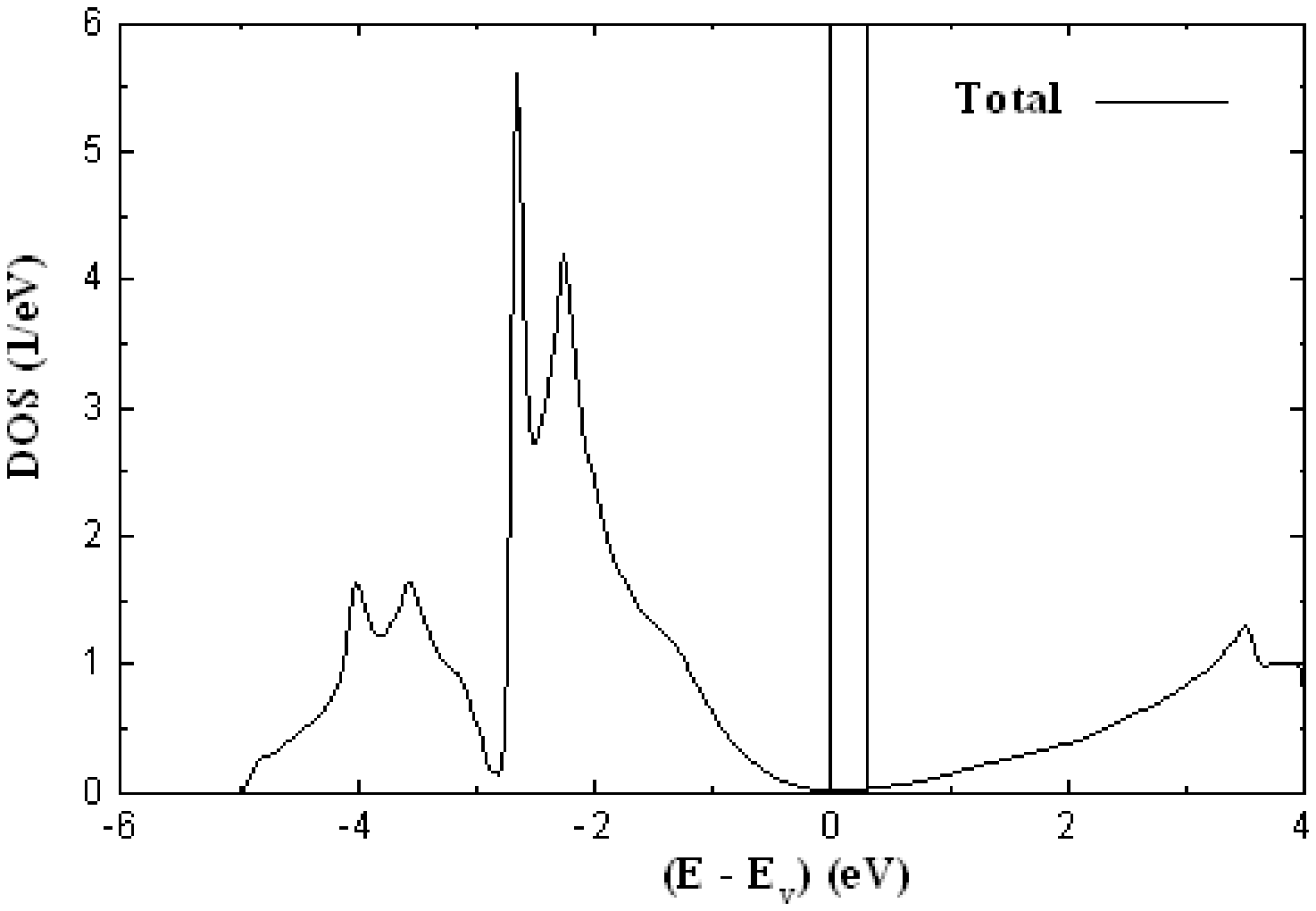}{b)}
\caption{Density of states of ScN at theoretical equilibrium: a) Total and site projected DOS with ASW calculations, b) Total DOS from FP-LAPW calculations.}
\label{fig2}
\end{center}
\end{figure}

Our band structure results are confirmed by the plots of density of states (DOS). Fig. \ref{fig2}-a shows the total and site projected density of states (PDOS) where the origin of energy is fixed at the maximum of VB.  Nitrogen  states are twice more intense than the Sc-{\it 3d} states which are mainly empty and are found prevailing within the conduction band. They are  separated by the energy gap. As a fact, Sc is at the beginning of the 3d transition metals period and $3d$ states are mainly found above E$_F$ as shown in fig. \ref{fig1}-b.  From this only Sc itinerant states, $s,p$-like, within the VB are found responsible of the mixing with N to ensure for the chemical bonding. This illustrates the chemical picture of Sc$^{\delta +}$N$^{\delta -}$ with $\delta \sim$3, meaning that the trivalent character of Sc is almost complete. 

Using FP-LAPW within Wien2k-method \cite{wien} calculations with the same GGA functional for ScN at experimental lattice constant, we also get the DOS. The calculations were carried with a k point integration within the irreducible wedge of the BZ up 85 k points (2200 total k points) and a muffin-tin ratio r$_{Sc}$/r$_N$=1,126. The DOS are shown in fig. 2-b in a narrow energy window around the top of the VB. The overall shape and skyline of the VB DOS is very similar to the PDOS in fig. 2-a obtained by ASW calculations. A band gap of $\sim$ 0.4 eV is identified and followed by a large dispersion of DOS as in fig. \ref{fig2}-b. The fact that in previous study \cite{Stampfl} a zero gap was obtained with FLAPW calculations could be due to the different choice of the muffin-tin radii of Sc and N (r$_{Sc}$/r$_N$=1,7). As it is not the main pupose of our present investigation, the reader is referred to the work of Qteish et al. \cite{Qteish} 
for more details about ScN and its electronic band structure.


\section{Magnetic properties and electronic structure for Sc$_{0.75}$Mn$_{0.25}$N}
\subsection{Structural stability and magnetic moment analysis}
Based on the recent experimental Sc$_{(1-x)}$Mn$_{x}$N results \cite{Albri}, where a substitution up to x=0.25 is obtained, we have investigated the stability and magnetic properties of Sc$_{0.75}$Mn$_{0.25}$N system. The results of this composition are obtained from calculations performed on Sc$_{3}$Mn$_{1}$N$_{4}$ simple cubic cell in a face-centred lattice (drived from rocksalt ScN with replacing one of the four Sc atoms in the cubic cell by Mn one). In a collective electrons scheme which is that of itinerant magnetism, such as the one used here, the magnetization arises from interband spin polarization, i.e., it is mediated by the electron gas. The theoretical equilibrium volume and the structural preference of Sc$_{0.75}$Mn$_{0.25}$N are obtained by calculating the variation of the total energy versus the volume in a cubic rocksalt (RS) structure and in the experimental face centred tetragonal-rocksalt ({\it fct}-RS) one \cite{Albri}. The instability toward magnetism in each structure can be obtained by comparing the spin-polarized (SP) and the non spin-polarized (NSP) total energy values at theoretical equilibrum volume. 


The results for theoretical lattice constants, total energy and other equilibrium properties are summarized in table II for the two studied structures together with the experimental data for the lattice constants.
At self consistency little charge transfer could be observed between atomic species, mainly from Sc and Mn towards N and ES; the presence of the latter allows ensuring for the covalence of the lattice by receiving small amounts of charge from N. It will be shown that the site projected DOS provide a better description of the quantum mixing between valence states.

The ferromagnetic state is largely preferred over the nonmagnetic one in the two structures.  Our results from ASW computations (see table II) indicate that, like the experimental finding \cite{Albri}, the ground state of Sc$_{0.75}$Mn$_{0.25}$N has an fct-RS structure rather than RS one with a better agreement with respective experimental lattice constants

For the former (fct-RS), the equilibrium volume and thus the lattice constants (a and c) were obtained by optimizing the c/a ratio in a fixed volume and then by repeating for different volumes from total energy calculations. The experimental lattice constants of Sc$_{1-x}$Mn$_{x}$N are found to decrease, as a function of Mn substitution, from ScN to {\it $\theta$}-MnN \cite{Albri}. Note that the difference in atomic covalent radii of Sc (1.44 \AA) and Mn (1.39 \AA) may allow explaining the observed trend. We found an in-plane {\it a} parameter in good  agreement with experiment, whereas our out-of-plane one {\it c} is somewhat smaller than the reported one. 
The structural stability of $fct$-RS with respect RS has been further checked using geometry optimization with GGA pseudo-potentials within the VASP method \cite{vasp}. Relaxation to zero strains confirmed through small energy differences that the tetragonal distortion RS $\rightarrow$ $fct$-RS is energetically favored.

For a possible instability towards  an antiferromagnetic arrangement (AF), total energy calculations of AF along [001] direction indicate that this AF order is less stable than the ferromagnetic one. Structural distortions in Mn-containing compounds are often related to some antiferromagnetic order. But as we have shown for the AF-[001] order, this should not be the reason here because the ferromagnetic state is the most stable; at least for such low amounts of Mn. However the presence of manganese in the trivalent matrix of ScN is likely to  induce a nearly trivalent behavior of Mn (d$^4$) which is a Jahn-Teller ion with  a closely t$_{2g}^3$e$_g^1$ in the octahedral field of nitrogen. This argument could be in favor of the observed tetragonal distortion of the lattice. 

\begin{figure}[h]
\begin{center}
\includegraphics[width=0.65\linewidth]{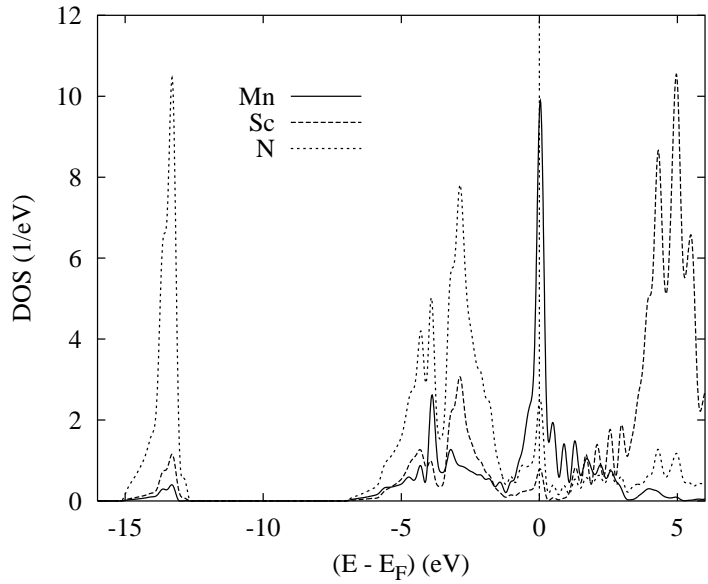}
\caption{Non spin polarized rocksalt Sc$_{0.75}$Mn$_{0.25}$N: site projected DOS at theoretical equilibrium}
\label{fig3}
\end{center}
\end{figure}

The non-spin polarized DOS of Sc$_{0.75}$Mn$_{0.25}$N are shown in Fig.  \ref{fig3}. In this plot as well as in next ones, the energy reference is with respect to the Fermi level. The VB shows large contribution from nitrogen states, mainly $s$-like at -14 eV and $p$-like around -5 eV. The latter are found to mix with itinerant states from both Mn and Sc. But the prevailing feature of the DOS is the large intensity at $E_F$ mainly arising from Mn-d states. This indicates an instability towards spin polarization as it can be inferred from the Stoner's theory \cite{Stoner} analysis. Within the Stoner theory of band ferromagnetism the large DOS at the Fermi level is related to the instability of the non magnetic state with respect to the onset of intraband spin polarization when n(E$_{F}$).I$>$1. In this so called Stoner criterion I is the Stoner integral which, for  Mn, is close to 0.43 eV \cite{janak}.  With n$_{Mn}$(E$_{F}$)= 9.9 eV$^{-1}$, the n(E{$_{F}$}).I value of 4.26 points to an unstable non magnetic state so that a large magnetic moment on the Mn sites should occur when two spin populations are accounted for in the calculations. As a consequence, the obtained ferromagnetic ground state is consistent with the Stoner theory. 
\subsection{Spin polarized electronic structure and chemical bonding}
\begin{figure}[h]
\begin{center}
\includegraphics[width=0.65\linewidth]{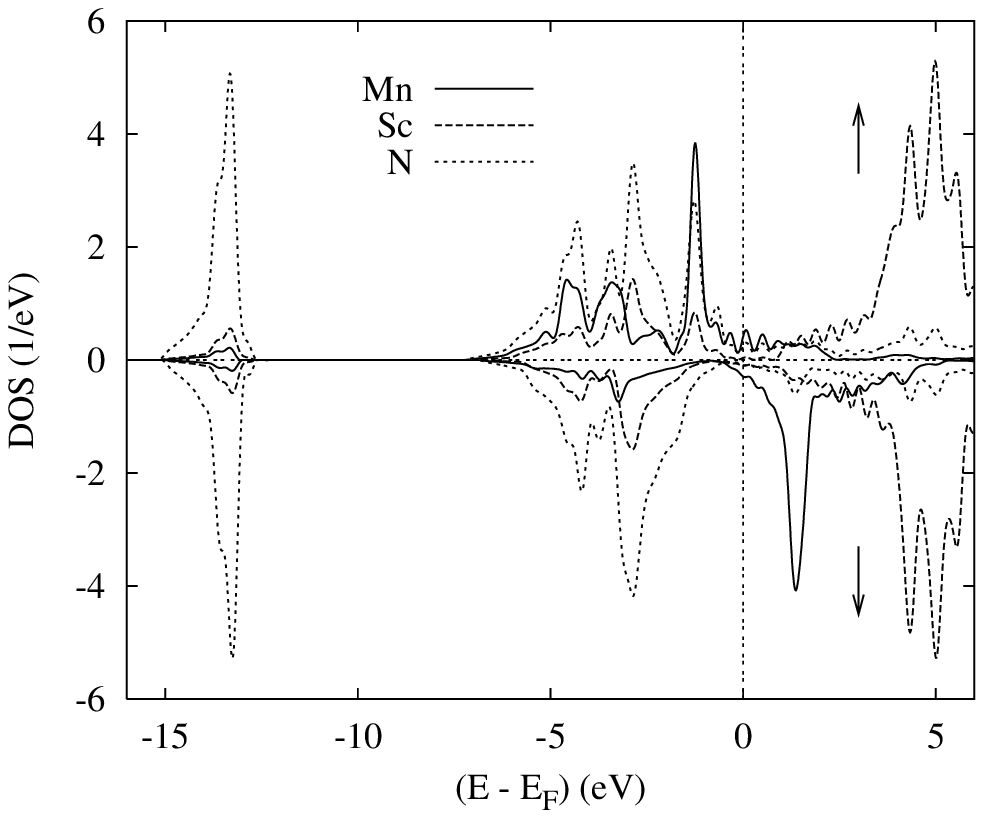}{a)}
\includegraphics[width=0.65\linewidth]{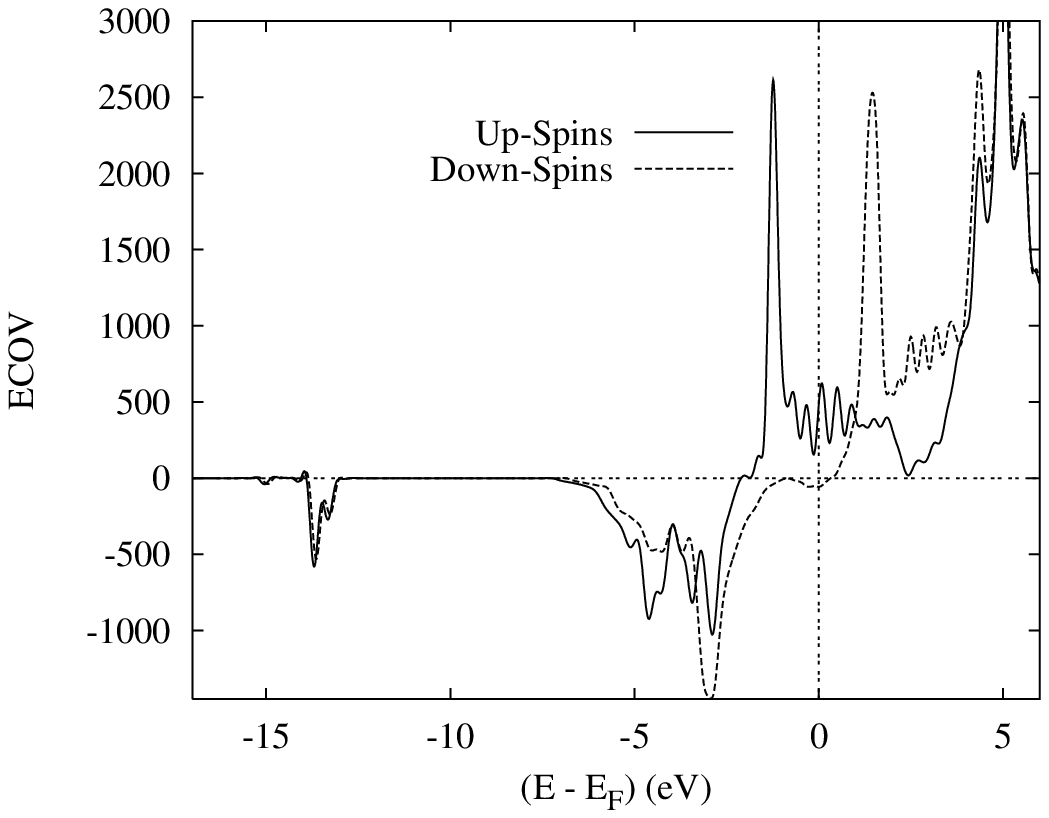}{b)}
\caption{Spin polarized fct-rocksalt Sc$_{0.75}$Mn$_{0.25}$N electronic structure at theoretical equilibrium: \\
  a) Partial density of states  (PDOS) and b) Total spin-up and spin-up ECOV.}
  \label{fig4}
\end{center}
\end{figure}

 The magnetic and electronic structures are illustrated in Fig.  \ref{fig4}, where the spin polarized densities of states (PDOS) and chemical bonding (ECOV) are plotted, for fct-RS-Sc$_{0.75}$Mn$_{0.25}$N alloy and indicating  a metallic character. (Those corresponding to the rocksalt structure are very similar and they are omitted here).  From Fig.  \ref{fig4}-a, a large exchange splitting can be noted for Mn atoms, with a nearly rigid band shift between the majority spin ($\uparrow$) to lower energy and  the minority spin ($\downarrow$) to higher energy due to the gain of energy from exchange. This confirms the high magnetic moment value of 3.44$\mu_B$ calculated for Mn atom. For Sc and N species, however, this splitting is almost absent (especially for the former) which agrees again with the obtained values (0.01 and 0.04$\mu_B$ respectively). In this context, we note that when the amount of Mn substitution is increased in Sc$_{1-x}$Mn$_{x}$N from x=0.25 to x=0.75, at the same volume,  complementary calculations  show that the Mn-moment decreases from 3.44 to 2.99 $\mu_B$ respectively. This is due to the increased overlapping of the Mn-d orbitals when the number of Mn atoms is raised. Unfortunately, in as far as no experimental magnetic data are available, our present  results could not be confronted with experiment. Nevertheless, in  supercell calculations (1 Mn atom in 64 ScN supercell), a magnetic moment of 2-3 $\mu_B$ was obtained \cite{Herwad}. The same figure shows that a large hybridization within the valence band, around -3 eV, exists between the {\it 3d} orbitals of the Mn and itinerant $s,p$ states of Sc on one hand  and the {\it 2p} ones of nitrogen on the other hand. The important feature which should be noted in the Sc$_{0.75}$Mn$_{0.25}$N DOS, is its intensity in the vicinity of the Fermi level. Although they are not completely vanishing, the Sc and N densities are very low from -0.5 eV to 1 eV and there is nearly a gap of 1.5 eV were it not for the presence of Mn states. However, Mn DOS are much larger in the vicinity of E$_F$. With respect to pure ScN, it seems here that the Sc and N densities  around the Fermi level are nearly not affected by Mn substitution. From DOS curves, it can be concluded that the major consequence of Mn-substitution in semiconducting ScN is to generate Mn-states situated in the gap which makes the compound metallic. 

For a better check of the chemical bonding within the nitride alloy, we show in figure  \ref{fig4}-b the ECOV curves representing the spin ($\uparrow$) and spin  ($\downarrow$) populations for all atoms. In this plot, the common feature for the two spin channels is their  antibonding character at positive energies (conduction band). At the valence band, however, a difference exists. Where the the spin ($\downarrow$) interactions are entirely bonding, a sharp antibondng peak is
 located just below the E$_F$ level for the spin ($\uparrow$) ones. This indicates roughly that the chemical bonds in
 Sc$_{0.75}$Mn$_{0.25}$N are spin resolved whereby the system is more bonded within minority spin ($\downarrow$) states than within majority spin ones. These chemical bonding features agree well with the densities of states analysis.   
           
\section{Summary} 
The main goal of the present work was to explore the properties of the recently grown Mn-substituted Sc$_{1-x}$Mn$_{x}$N with x=0.25  ternary alloys from first-principles. Using the self-consistent DFT-based ASW method, the electronic properties of the scandium mononitride (ScN), which was recently found semicondctor and not semimetal as was suggested early, are firstly reviewed. Structural, electronic and magnetic properties of Sc$_{0.75}$Mn$_{0.25}$N compound are then investigated. The calculated total-energy versus volume for ScN, in its known rocksalt structure, lead to a theoretical lattice parameter of {\it a}=4.5 \AA ~in perfect agreement with experiments. Like the recent experimental and theoretical reports, we have found ScN as an indirect $\Gamma$-X band gap semiconductor, but our calculated band width (0.54 eV) is lower than the (0.9 eV) experimentally suggested one. In view of the available theoretical reports, our main conclusion is that the use of an accurate approximation of the exchange-correlations potential is crucial to predict the nature and the width of ScN gap. Concerning the Sc$_{0.75}$Mn$_{0.25}$N nitride, it was considred in both: the perfect rocksalt structure and the experimentally indexed face centred tetragonal-rocksalt (fct-rocksalt) one. The structural results confrim the stability of the latter over the former, and a very good agreement is found for the lattice constants  where the calculations reproduce pefectly the experimental in-plane {\it a} lattice  constant.The out-of-plane {\it c} value is slightly lower than experiments. Magnetic moments of 3.44, 0.01 and 0.04$(\mu_B)$ were obtained for Mn, Sc and N atoms respectively, and confirmed by the DOS plots which agree well with the rigid-band model of magnetism. With respect to  pristine ScN, our densities of states and chemical bonding analysis show that the Mn-incorporation in ScN structure is responsible of the metallic character (by introducing Mn-states at E$_F$ level). It  also indicates that this does not affect significantly the original ScN gap, where the Sc and N DOS intensities around E$_F$ remain very low. The chemical bonding in Sc$_{0.75}$Mn$_{0.25}$N is found to be spin resolved with a more bonded character with spin ($\downarrow$) population than spin ($\uparrow$) one.       
\section{Acknowledgments}
We acknowledge computational facilities provided by the M3PEC-M\'esocentre of the University Bordeaux 1 (http://www.m3pec.u-bordeaux1.fr), financed by the ``Conseil R\'egional d'Aquitaine'' and the French Ministry of Research and Tecnology. A.H. thanks the University of Bejaia, Algeria for a grant. 

{}

\newpage
\begin{table}
\caption{\label{tab:table1}Lattice constant and Band gaps in rocksalt ScN.}
\begin{ruledtabular}
\begin{tabular}{llllccc} 
Method                                          &lattice       &Band \\
                                                &constants(\AA)&gap (eV)\\              
 \hline
Present ASW-GGA                                 & 4.50         &(indirect $\Gamma$-X) 0.54\\
FLAPW-GGA\footnote{Ref. \cite{Stampfl}}         &      4.50    & 0.0                       \\
FLAPW-SX-LDA\footnote{Ref. \cite{Stampfl}}      &      4.50    & (indirect $\Gamma$-X) 1.58 \\
LMTO-GW\footnote{Ref. \cite{Lamb}}              &              & (indirect $\Gamma$-X) 0.9 \\
PP-PW-cLDA-GW\footnote{Ref. \cite{Qteish}}      &              & (indirect $\Gamma$-X) 0.9 \\
PP-PW-GGA\footnote{Ref. \cite{Qteish}}          &     4.53     & (indirect $\Gamma$-X) -0.03 \\
Expt.\footnote{Ref. \cite{Schilf,Gubanov}}      &     4.50     &    \\
Expt.\footnote{Ref. \cite{Albrithen}}           &              & (indirect $\Gamma$-X) 0.9   \\
\end{tabular}
\end{ruledtabular}
\end{table}
\begin{table}
\caption{\label{tab:table2}Calculated equilibrium properties for Sc$_{0.75}$Mn$_{0.25}$N in rocksalt (RS)and face centred
 tetragonal-rocksalt ({\it fct}-RS) }
\begin{ruledtabular}
\begin{tabular}{llllccc}
equilibrium             & lattice constants (\AA)    & total energy(eV) &$m (\mu_B)$\\
properties              &                            &                  &           \\
\hline
                        &                            &      RS          &          \\
NSP                     &a=4.393                     & -24961.224       &          \\
SP                      & a=4.437                    & -24961.554                 &Mn=3.50   \\
                     &                              &                             &Sc=0.01   \\
                     &                              &                             &N=0.04    \\
\hline
                     &                              & fct-RS                      &          \\
 NSP                 & a=4.417                      & -24961.344                  &          \\
                     & c=4.311                      &                                        \\
 SP-FM               & a=4.443                      & -24961.687                  &Mn=3.44   \\
                     & c=4.337                      &                             &Sc=0.01   \\
                     &                              &                             &N=0.04    \\
 SP-AFM              & a=4.443                                 & -24961.626                &Mn=3.43   \\
                     &   c=4.337                           &                             &Sc=0.01   \\
                     &                              &                             &N=0.03  \\
Expt.\footnote{Ref. \cite{Albri}}  & a=4.440        &                             &        \\
                                   & c=4.405        &                             &
\end{tabular}
\end{ruledtabular}
\end{table}

\end{document}